\documentclass[11pt,a4paper]{article}
\usepackage[hyperref]{emnlp2020}
\usepackage{times}
\usepackage{latexsym}
\usepackage{amsmath,amsthm,amssymb}
\usepackage{graphicx}
\usepackage{enumitem}
\usepackage{setspace}
\usepackage{color}
\usepackage[normalem]{ulem} 
\usepackage{xcolor}
\usepackage{subfigure} 
\theoremstyle{definition}

\usepackage{multirow}
\usepackage{subfigure} 

\usepackage{ textcomp }
\usepackage{ gensymb }
\usepackage{ wasysym }
\usepackage[many]{tcolorbox}
\usepackage{tikz}

\usepackage{booktabs} 
\usepackage{microtype}
\usepackage[table]{colortbl}

\aclfinalcopy 
\def\aclpaperid{872} 

\newcommand{\squishlist}{
	\begin{list}{$\bullet$}
		{ \setlength{\itemsep}{0pt}
			\setlength{\parsep}{3pt}
			\setlength{\topsep}{3pt}
			\setlength{\partopsep}{0pt}
			\setlength{\leftmargin}{1.5em}
			\setlength{\labelwidth}{1em}
			\setlength{\labelsep}{0.5em} } }
	
	\newcommand{\squishlisttwo}{
		\begin{list}{$\bullet$}
			{ \setlength{\itemsep}{0pt}
				\setlength{\parsep}{0pt}
				\setlength{\topsep}{0pt}
				\setlength{\partopsep}{0pt}
				\setlength{\leftmargin}{2em}
				\setlength{\labelwidth}{1.5em}
				\setlength{\labelsep}{0.5em} } }
		
		\newcommand{\squishend}{
	\end{list}  }
\title{Where Are the Facts? Searching for Fact-checked Information \\ to Alleviate the Spread of Fake News}

\author{Nguyen Vo {\normalfont{and}} Kyumin Lee\\
	Computer Science Department, Worcester Polytechnic Institute \\
	Worcester, Massachusetts, 01609, USA \\
	\texttt{\{nkvo,kmlee\}@wpi.edu} \\  
}

\begin{document}
\maketitle
\begin{abstract}

Although many fact-checking systems have been developed in academia and industry, fake news is still proliferating on social media. These systems mostly focus on fact-checking but usually neglect \textit{online users} who are the main drivers of the spread of misinformation.
How can we use fact-checked information to improve users’ consciousness of fake news to which they are exposed? How can we stop users from spreading fake news? To tackle these questions, we propose a novel framework to search for fact-checking articles, which address the content of an original tweet (that may contain misinformation) posted by online users. The search can directly warn fake news posters and online users (e.g. the posters' followers) about misinformation, discourage them from spreading fake news, and scale up verified content on social media. Our framework uses both text and images to search for fact-checking articles, and achieves promising results on real-world datasets. Our code and datasets are released at \url{https://github.com/nguyenvo09/EMNLP2020}.	
\end{abstract}

\section{Introduction}
The rampant spread of biased news, partisan stories, false claims and misleading information has raised heightened societal concerns in recent years. 
Many reports pointed out that fabricated stories possibly caused citizens' misperception about political candidates \cite{allcott2017social}, 
manipulated stock prices \cite{kogan2019fake} and threatened
public health \cite{Ashoka2020,WhatsAppIndia}.
\begin{figure}[t]
	\includegraphics[width=\linewidth, height=1.7in, trim={2.0cm 7.3cm 20.5cm 2.5cm},clip]{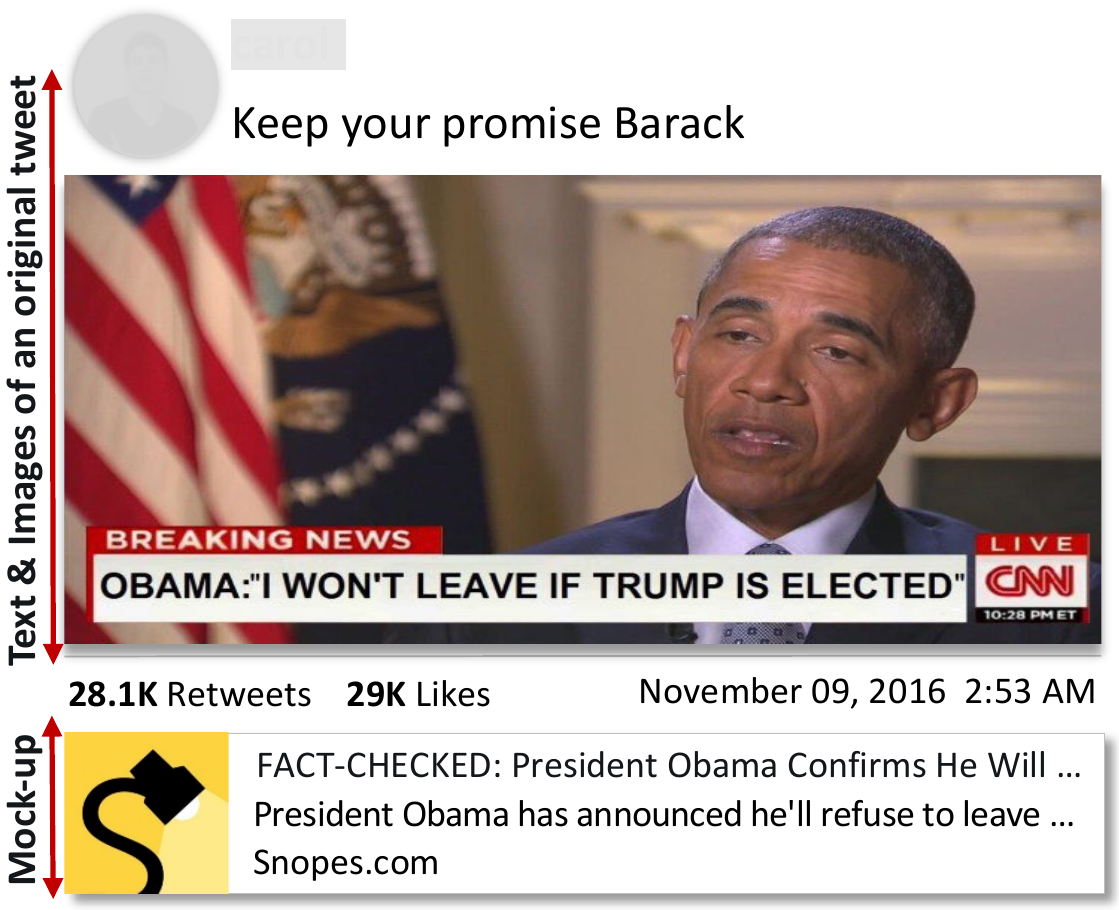}
	\caption{An original tweet and a mock-up of how a correctly retrieved fact-checking article is presented.}
	\vspace{-5pt}
	\label{fig:fake_news_example}
	\vspace{-15pt}
\end{figure}

The proliferation of misinformation has provoked the rise of fact-checking systems worldwide. Since 2014, the number of fact-checking outlets has totally increased 400\% in 60 countries \cite{poynter}.
However, fabricated stories and hoaxes are still pervading our cyberspace. Fig. \ref{fig:fake_news_example} shows an example of a fake quote related to Barack Obama. The quote had been debunked by Snopes \cite{ObamaSnopes} on September 08, 2016 but two months later, it appeared again inside an original tweet posted by a Twitter user (called an original poster) and was retweeted over 28 thousand times. Perhaps, the original poster and people who shared the original tweet did not know if it was fact-checked or they might share it simply because it was suitable for their personal preferences or ideologies \cite{lewandowsky2012misinformation}. In other words, existing fact-checking systems mainly focus on detection but neglect online users who play the critical role in spreading fake news. After detecting fake news, what are the next steps to discourage people from sharing it? Recent studies \cite{vo2018rise, vo2019learning} tried to curb the above weakness. However, these approaches are not proactive since they rely on fact-checkers who may be unreliable.

Recent works showed that when seeing fact-checked information, users' likelihood to delete fake news's shares went up 400\% \cite{friggeri2014rumor} and 95\% of the time users did not further consume or go through fake news \cite{FacebookMark95}. Observing the downside of existing methods and impacts of broadcasting verified news, our goal is to search for \textit{fact-checking articles} (FC-articles) which address the content of original tweets (i.e. confirming, supporting, debunking or refuting). We show a mock-up of how a relevant FC-article is linked/displayed given an original tweet in Fig. \ref{fig:fake_news_example}. 
By searching for FC-articles and incorporating fact-checked information into social media posts, we can warn users (e.g. followers of original posters) about fake news to which they are exposed. The search also proactively scales up volume of verified content on social media. 
However, achieving the goal is challenging since we need to solve two problems: \textbf{(P1)} what information in original tweets should we use to find correct FC-articles? and \textbf{(P2)} how can we design a framework to retrieve and rank FC-articles? 


With the first problem \textbf{(P1)}, we can use original tweets' text to find FC-articles. However, this approach is suboptimal since fake news can appear in many forms (e.g. text, images, videos) \cite{friggeri2014rumor,CNNDeppFake} as shown in Fig. \ref{fig:fake_news_example}. 
Thus, we propose to use both text and images of original tweets to search for FC-articles. Regarding the second problem \textbf{(P2)}, 
we propose a framework consisting of two key steps: (1) using a basic retrieval (i.e. BM25) to find initial lists of candidate FC-articles and then (2) re-ranking the initial lists by using advanced models for ranking refinement. In the first step, since original tweets' text may be insufficient to find correct articles as shown in Fig. \ref{fig:fake_news_example} where there is no meaningful information in the text but in the image, we propose to expand original tweets' text by using text inside original tweets' images. For the second step, we propose an attention mechanism to focus on key textual matching signals and jointly integrate them with visual information to boost ranking quality.
By tackling these issues, our contributions are as follows:
\squishlist
\item To the best of our knowledge, our study is the first one that searches for fact-checking articles to increase users' awareness of fact-checked information when they are exposed to fake news. 
\item We propose a novel neural ranking model which jointly utilizes textual and visual matching signals. The model is also integrated with a novel attention mechanism. 
\item Experiments on two datasets demonstrate effectiveness and generality of our model over state-of-the-art retrieval techniques. 
\squishend

\section{Related Work}

\noindent\textbf{Fake News and Fact-checking.}
Fake news detection methods mainly use linguistics and textual content \cite{zellers2019defending,zhao2015enquiring,wang2017liar,shu2019defend}, temporal spreading patterns \cite{liu2018early,ma2018rumor}, network structures \cite{wu2018tracing,liu2020kernel} and users' feedbacks \cite{vo2019learning,vo2020standing,shu2019defend}. Studies about multimodal fake news detection \cite{gupta2013faking,wang2018eann} are different from ours since their inputs are text and images of tweets while our inputs are pairs of a multimodal tweet and a FC-article.



Our work is closely related to evidence-aware fact-checking. \citet{thorne2018fever,nie2019combining} built a pipeline to find documents and sentences to fact-check mutated claims generated from Wikipedia pages, \citet{wang2018relevant} aimed to find webpages related to given FC-articles and predict their stances on claims in the FC-articles. \citet{popat2018declare} only focused on fact-checking and \cite{shaar2020known} detected previously fact-checked claims. Our paper deviates from these work since we aim to find FC-articles given multimodal fake news in social media posts. As our goal is to increase users' awareness of verified news, studies about fact-checkers \cite{vo2018rise,vo2019learning,you2019attributed} are close to ours.

\noindent\textbf{Neural Ranking Models for Text Search.}
Neural ranking models for text search mainly fall into two groups: semantic matching and relevance matching models. The former one seeks to learn representations of a query and a document, and measure their similarity \cite{huang2013learning,shen2014latent,severyn2015learning,nie2019combining,zhu2019hierarchical}, 
while the later one \cite{chen2018mix,hui2017pacrr,pang2016text,guo2016deep,xiong2017end,hui2018co,dai2018convolutional} aims to capture relevant matching signals between a query and a document based on word interactions. There are methods unifying two categories such as \citet{mitra2017learning,rao2019bridging}. 
Our model can be viewed as a relevance matching method in which a novel attention mechanism is designed to focus on crucial word interactions. 

\noindent\textbf{Neural Models for Multimodal Retrieval.}
Multimodal data (e.g. text and images) are used in
cross-modal retrieval \cite{cao2016deep,wang2017adversarial,balaneshin2018deep,chen2016context}, visual Q\&A task \cite{kim2018bilinear}, product search \cite{laenen2018web, guo2018multi} and so on. Our work is the first using multimodal data in social media posts to search for verified information.




\section{Our framework}
Given an original tweet 
$q$ and a FC-article 
$d$, where every original tweet $q$ contains text and images and the article $d$ contains text and/or images, we aim to derive function $f(q,d)$ which determines their relevancy\footnote{\scriptsize{Relevance means that the fact-checking article fact-checks the query}}. We use $f(q,d)$ to rank all FC-articles.

\begin{figure*}[t]
	\includegraphics[width=1\textwidth, height=3.0in,trim={0.1cm 1.0cm 0.4cm 0.1cm},clip]{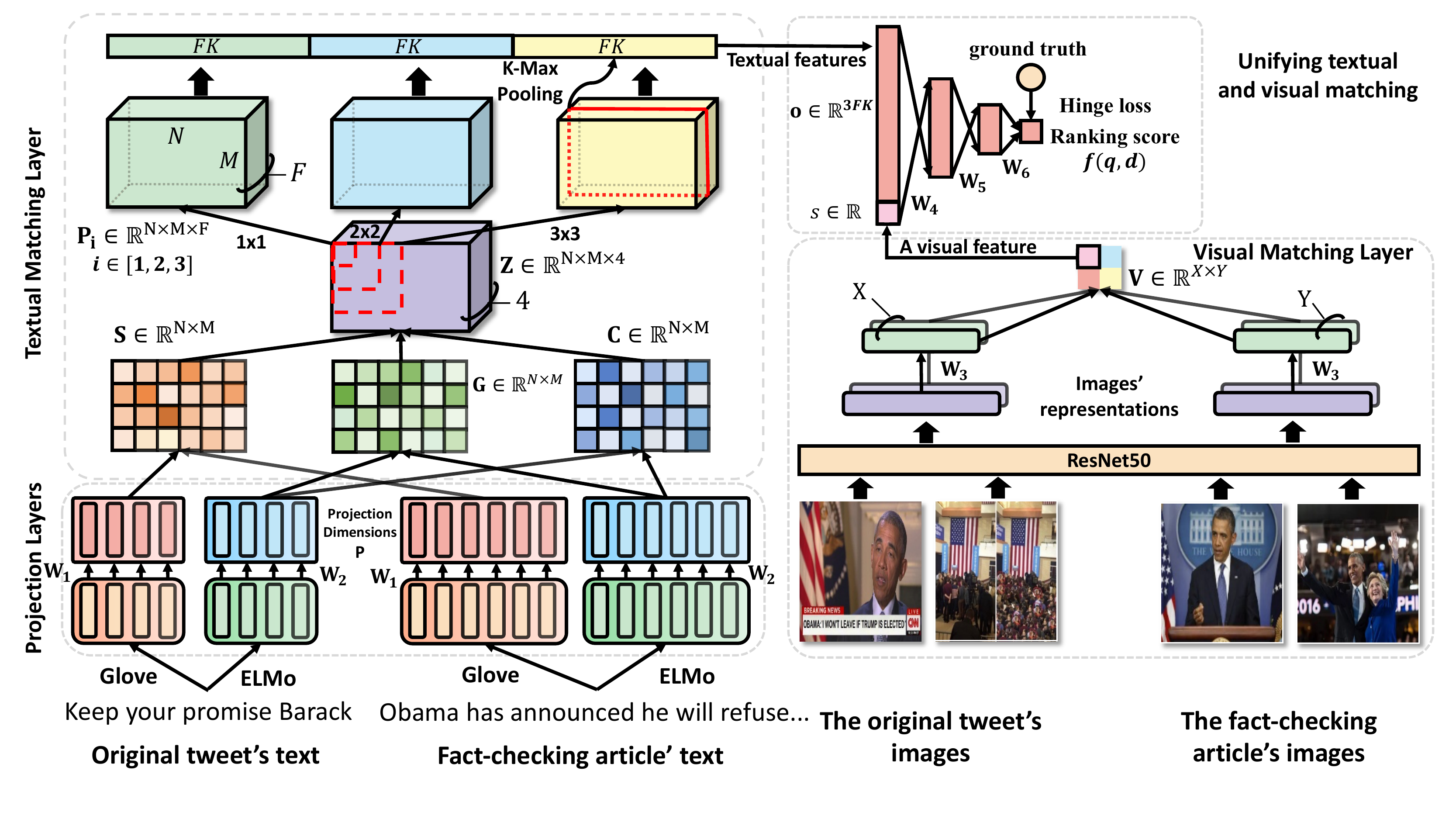}
	\caption{Our proposed model MAN}
	\vspace{-5pt}
	\label{fig:novel_model1}
	\vspace{-10pt}
\end{figure*}

Following \cite{thorne2018fever}, we adopt the re-ranking methodology as follows: (1) quickly retrieving candidate FC-articles/documents\footnote{\scriptsize{We use fact-checking articles, articles and documents interchangeably}} for each original tweet/query\footnote{\scriptsize{We use original tweets and queries interchangeably}} by a basic retrieval and (2) re-ranking the candidates by our MAN (\underline{M}ultimodal \underline{A}ttention \underline{N}etwork) as shown in Fig.~\ref{fig:novel_model1}. 
We describe our input representations, basic retrieval and MAN in following subsections.


\subsection{Input Representations}
We denote text and images of an original tweet $q$ as $(q_{text}, q_{images})$ where $q_{text}$ is a sequence of $N$ words $\{w^q_i\}_{i=1}^N$ and $q_{images}$ is a list of $X$ images $\{v^q_i\}_{i=1}^X$. Similarly, text and images of a fact-checking article $d$ are denoted as $(d_{text}, d_{images})$ where $d_{text}$ is a sequence of $M$ words $\{w^d_j\}_{j=1}^M$ and $d_{images}$ is a list of $Y$ images $\{v^d_j\}_{j=1}^Y$.

\subsection{Basic Retrieval}
We use BM25 as a basic retrieval due to its good performance compared with several ranking models \cite{mcdonald2018deep,pang2017deeprank}. Since using tweets' text may be insufficient to find relevant articles, we expand queries' text by using text extracted from images. For example, in Fig.~\ref{fig:fake_news_example}, text extracted from the image is \textit{Breaking News: Obama: "I won't leave if Trump is elected"}. Following \cite{vosoughi2018spread}, we use a tool \cite{ocrspace} to extract text in images. 
To our knowledge, our work is the first one using text in images to find verified information. 
\subsection{Multimodal Attention Network (MAN)}
MAN has four components: (1) projection layers, (2) textual matching layer (3) visual matching layer and (4) unifying textual and visual information. 
\subsubsection{Projection Layers}
We use two projection layers: one for Glove embeddings and the other one for contextual word embeddings. 
\setlength{\abovedisplayskip}{4pt}
\setlength{\belowdisplayskip}{4pt}

\noindent\textbf{Projection layer for Glove embeddings.}
\label{sec:projection_layer_word_embeddings}
Each word $w$, which can be $w^q_i$ or $w^d_j$, is mapped into a vector $\textbf{t}\in \mathbb{R}^{300}$ by a fixed word embedding layer initialized by Glove embeddings \cite{pennington2014glove}. Then, the vector $\textbf{t}$ is projected into $\textbf{g}\in \mathbb{R}^P$ by a trainable linear layer shown in Eq. \ref{eq:map_word_embeddings}.
\begin{equation}
	\textbf{g} = tanh(\textbf{W}_1 \cdot \textbf{t} + \textbf{b}_1)
	\label{eq:map_word_embeddings}
\end{equation}
where $\textbf{W}_1\in \mathbb{R}^{P\times {300}}$, $\textbf{b}_1\in \mathbb{R}^P$. $P$ is projection dimensions. 
After going through the linear layer, we denote $\textbf{g}^q_i \in \mathbb{R}^P$ and  $\textbf{g}_j^d\in \mathbb{R}^P$ as representations of word $w^q_i$ and word $w^d_j$, respectively. 

\noindent\textbf{Projection layer for contextual word embeddings.}
Since Glove embeddings do not reflect context of words in queries and articles, 
we integrate ELMo \cite{peters2018deep} as a static encoder to generate contextual word embeddings. 
ELMo maps each word $w$, which can be $w^q_i$ or $w^d_j$, into a vector $\boldsymbol{\ell}\in \mathbb{R}^{1024}$ which is then projected into $\textbf{h} \in \mathbb{R}^{P}$ by a trainable linear layer shown in Eq.~\ref{eq:map_elmo}.
\begin{equation}
	\textbf{h} = tanh(\textbf{W}_2 \cdot \boldsymbol{\ell}+ \textbf{b}_2)
	\label{eq:map_elmo}
\end{equation}
where $\textbf{W}_2\in \mathbb{R}^{P\times 1024}$, $\textbf{b}_2\in \mathbb{R}^P$. $P$ is projection dimensions. 
After going through the linear layer, we denote $\textbf{h}^q_i \in \mathbb{R}^P$ and  $\textbf{h}_j^d\in \mathbb{R}^P$ as contextual representations of words $w^q_i$ and $w^d_j$, respectively.
\subsubsection{Textual Matching Layer}
\label{sec:text_matching_layer}
We derive (1) Glove embeddings interactions, (2) attended interaction matrix and (3) contextual word embedding interactions, and input them to convolution neural networks (CNNs) for feature extraction. 

\noindent\textbf{Glove Embeddings Interactions.} An article may be relevant to an original tweet if they have overlapping words or similar words. To capture such signals, we use cosine similarity to derive matrix $\textbf{S} \in \mathbb{R}^{N\times M}$ as shown in Eq.~\ref{eq:cosine_sime_S}.   
\begin{equation}
\textbf{S}_{ij} = {{  {\textbf{g}^q_i}^T\cdot\textbf{g}^d_j} \over {||\textbf{g}^q_i||\times ||\textbf{g}^d_j||}}, \text{i = 1..N, j = 1..M}
\label{eq:cosine_sime_S}
\end{equation}

Let's look at an example of matrix $\textbf{S}$ in Fig.~\ref{fig:glove_cosine_sim_mat} where x-axis is an article and y-axis is a query. Roughly speaking, matrix $\textbf{S}$ looks like a \textit{gray-scale image} in which the overlapping phrase `at a costume party' is like a \textit{segment} at the bottom of the image, suggesting the article is relevant to the query. To capture such patterns, CNNs are widely used. 


\noindent\textbf{Attended Interaction Matrix. }
Matrix $\textbf{S}$ captures overlapping words between a query and an article. However, when word $w^q_i$ is same as word $w^d_j$, sometime they may not have the same meaning. Thus, we need an attention mechanism to avoid over-reliance on raw similarities in matrix $\textbf{S}$. Inspired by \citet{tay2019compositional}, we measure how dissimilar $w^q_i$ and $w^d_j$ are based on Euclidean distance between their contextual representations as follows: 
\begin{equation}
\textbf{G}_{ij} = 2 \times \boldsymbol{\sigma}(-||\textbf{h}^q_i - \textbf{h}^d_j||),\text{i = 1..N, j = 1..M}
\label{eq:disim_cosine}
\end{equation}
where $\boldsymbol{\sigma}(.)$ is a sigmoid function. Since Euclidean distance is non negative, $\boldsymbol{\sigma}(-||\textbf{h}^q_i - \textbf{h}^d_j||)$ will be in $(0, 0.5]$ and $\textbf{G}_{ij}$ will be in $(0, 1]$. Therefore, we can use $\textbf{G}_{ij}$ to attend to $\textbf{S}_{ij}$ as follows:
\begin{equation}
	\textbf{A}_{ij} = \textbf{S}_{ij} \times \textbf{G}_{ij}, \text{i = 1..N, j = 1..M}
	\label{eq:attended_matrix}
\end{equation}
It is clear to see that when the distance between $\textbf{h}^q_i$ and $\textbf{h}^d_j$ is large, $\textbf{G}_{ij}$ will be closer to 0 which helps downgrade impact of $\textbf{S}_{ij}$. From Eq.~\ref{eq:attended_matrix}, we can form attended interaction matrix $\textbf{A} \in \mathbb{R}^{N\times M}$. 

To our knowledge, our work is the first one using dissimilarity between contextual word embeddings to attend to interactions of Glove embeddings.




\noindent\textbf{Contextual Word Embeddings Interactions.}
In our case studies in Section \ref{sec:case_studies}, we find that contextual word embeddings are able to capture high similarity between a typo and a normal word (e.g. {hillar} vs. {hillary}) while Glove embeddings fail to do so. To further exploit contextual embeddings, we derive matrix $\textbf{C} \in \mathbb{R}^{N\times M}$ as follows: 
\begin{equation}
\textbf{C}_{ij} = {{  {\textbf{h}^q_i}^T\cdot\textbf{h}^d_j} \over {||\textbf{h}^q_i||\times ||\textbf{h}^d_j||}}, \text{i = 1..N, j = 1..M}
\label{eq:context_dot}
\end{equation}
Again, we can view matrix $\textbf{C}$ as a greyscale image as shown in Fig.~\ref{fig:elmo_cosine_sim_mat}. In addition to cosine similarities, we found that using bilinear function \cite{rao2019bridging} works pretty well as well.

\noindent\textbf{Textual Feature Extraction.} 
We stack matrices \textbf{S} (Eq.~\ref{eq:cosine_sime_S}), \textbf{A} (Eq.~\ref{eq:attended_matrix}) and \textbf{C} (Eq.~\ref{eq:context_dot}) and $\textbf{S} - \textbf{C}$ to generate a tensor $\textbf{Z} \in \mathbb{R}^{N\times M\times 4}$ shown in Eq.~\ref{eq:dissim_tensor_attention}. The matrix $\textbf{S} - \textbf{C}$ is used to make our model aware of differences between interaction matrices.
\begin{equation}
\textbf{Z} = [\textbf{S}\oplus\textbf{A}\oplus\textbf{C}\oplus(\textbf{S}-\textbf{C})]
\label{eq:dissim_tensor_attention}
\end{equation}
`$\oplus$' denotes matrix stacking. 
We apply $n$ CNNs on tensor \textbf{Z} to extract features. The $i^{th}$ CNN is performed with kernel size, stride and the number of filters equal to $i\times i\times 4$, 1 and $F$, respectively. 
The output feature map of the $i^{th}$ CNN layer is $\textbf{P}_i\in \mathbb{R}^{N\times M\times F}$, $i\in\{1..n\}$. Note, padding zeros are used to ensure $\textbf{P}_i$ has size of $N\times M\times F$.

Next, we apply k-max pooling on each $j^{th}$ output channel of $\textbf{P}_i$ denoted as $\textbf{P}_i[:\,,:\,,j]\in \mathbb{R}^{N\times M}$ to generate vector $\textbf{o}_{i,j} \in \mathbb{R}^{K}$ as shown in Eq.~\ref{eq:kmax_pool}.
\begin{equation}
\textbf{o}_{i,j} = \text{kmax}(\textbf{P}_i[:\,,:\,,j]),\,\text{i = 1..n, j = 1..F}
\label{eq:kmax_pool}
\end{equation}

Finally, nF vectors $\textbf{o}_{i,j}$ are concatenated to create textual features
vector $\textbf{o} \in \mathbb{R}^{nFK}$ shown in Eq.~\ref{eq:textual_features}.
\begin{equation}
	\textbf{o} = [\textbf{o}_{1,1};...;\textbf{o}_{i,j};...;\textbf{o}_{n,F}] \in \mathbb{R}^{nFK}
	\label{eq:textual_features}
\end{equation}
\subsection{Visual Matching Layer}
A fixed pretrained ResNet50 \cite{he2016deep} maps an image $v$, which is either an image of an original tweet $v^q_i$ or an image of a FC-article $v^d_j$, into vector $\textbf{v} \in \mathbb{R}^{H}$ which is then projected into vector $\textbf{m}\in \mathbb{R}^{T}$ by a trainable linear layer: $\textbf{m}= \textbf{W}_3 \cdot \textbf{v} + \textbf{b}_3$
, where $\textbf{W}_3 \in \mathbb{R}^{T\times H}$ and $\textbf{b}_3 \in \mathbb{R}^{T}$. $H$ and $T$ are set to 2048 and 300, respectively.
After the linear layer, we denote $\textbf{m}^q_i\in \mathbb{R}^{T}$ and $\textbf{m}^d_j\in \mathbb{R}^{T}$ as representations of $v^q_i$ and $v^d_j$, respectively.

Intuitively, an article is relevant to a query if the article has images similar to the query's images. Thus, we derive matrix $\textbf{V} \in \mathbb{R}^{X\times Y}$ of pairwise similarities of images in Eq.~\ref{eq:visual_sim}.
\begin{equation}
\textbf{V}_{ij} = {{  {\textbf{m}^q_i}^T\cdot\textbf{m}^d_j} \over {||\textbf{m}^q_i||\times ||\textbf{m}^d_j||}} ,\text{i = 1..X, j = 1..Y}
\label{eq:visual_sim}
\end{equation}
Similar to \cite{rao2019bridging,rao2019multi}, we pool the largest pairwise similarity $s$ as a visual feature as follows:
\begin{equation}
	s = max(\textbf{V}), \text{where}\, s\in \mathbb{R}
	\label{eq:max_visual_consine_sim}
\end{equation}
When the article has no images, $s$ is set to $-1$.

\subsection{Unifying Textual and Visual Information}
We unify textual and visual information by appending scalar $s$ (Eq.~\ref{eq:max_visual_consine_sim}) to vector $\textbf{o}$ (Eq.~\ref{eq:textual_features}), denoted as $[\textbf{o};s]$, and derive $f(q, d)$ as shown in Eq.~\ref{eq:cman_score}.
\begin{equation}
f(q, d) = \textbf{W}_6\cdot relu(\textbf{W}_5\cdot relu(\textbf{W}_4\cdot [\textbf{o};s]))
\label{eq:cman_score}
\end{equation}
where $\textbf{W}_4 \in \mathbb{R}^{128\times (nFK + 1)}$, $\textbf{W}_5\in \mathbb{R}^{64\times 128}$ and $\textbf{W}_6\in \mathbb{R}^{1\times 64}$. We remove biases to avoid clutter. Our model is trained on triples consisting of a query $q$, relevant document $d^+$ and non-relevant document $d^-$, minimizing hinge loss in Eq.~\ref{eq:hinge_loss}. 
\begin{equation}
\resizebox{0.866\linewidth}{!}
{$\mathcal{L}(q, d^+,d^-) = max(0, 1 - f(q, d^+) + f(q, d^-))$}
\label{eq:hinge_loss}
\end{equation}

\section{Data Collection}
\label{sec:datasets}
Finding FC-articles, which address an original tweet, is laborious since we have to read many FC-articles even when using search engines \cite{popat2017truth,popat2018declare}. To reduce labeling efforts, we looked at existing datasets \cite{jiang2018linguistic,vosoughi2018spread,vo2019learning} and found that a dataset in \citet{vo2019learning} met our need. The dataset provides non-anonymized pairs of an original tweet and its reply in which FC-articles, from two major fact-checking sites \textit{snopes.com} and \textit{politifact.com}, are embedded. \emph{Fact-checkers} in \citet{vo2019learning} replied to the original tweet posters with FC-articles as evidence. From the original tweets' replies, we generate pairs of an original tweet $q$ and a FC-article $d$. We also only kept original tweets where text and images are both available.

After preprocessing, we have 19,341 original tweet in English and FC-article pairs $(q, d)$ in which there are 18,961 unique original tweets and 2,845 FC-articles.
Following \citet{vosoughi2018spread}, a labeling step is conducted to ensure that in each pair, the article fact-checks the original tweet. We hired native U.S. English speakers since they were more likely to be familiar with topics in the tweets and FC-articles. The labelers labeled each pair $(q,d)$ as 1 if the article $d$ fact checked the tweet $q$. Otherwise, they labeled it as 0. They were trained directly by the authors and were asked to label several examples as exercises to ensure that they fully understood the task. We required labelers to read the original tweet's text, the article's text and images, and developed a labeling UI to help labelers to quickly explore the linked FC-articles shown in Fig.~\ref{fig:labeling_ui} in our appendix.
For each pair, three different labelers labeled it. The final label is based on the majority vote. The Kappa value is 0.56, suggesting moderate agreement among the labelers \cite{viera2005understanding}.

The moderate agreement between labelers was because there were many pairs of an original tweet and a FC-article where the tweet and the article are topically similar but the article does not fact-check the tweet. For example, the tweet is about Hillary Clinton’s mishandled classified emails while the article fact-checks if she gave uranium to Russia. Both the tweet and the article were about Hillary Clinton but the article did not precisely fact-check the tweet’s content. As we utilized the dataset in \cite{vo2019learning} which was collected during the 2016 U.S. presidential election, many tweets and FC-articles were about misinformation related to Hillary Clinton and Donald Trump, leading to topically similar pairs which might confuse labelers.
After labeling, we have a full dataset of 13,239 positive pairs made by 13,091 original tweets and 2,170 FC-articles.

We observe that there may be \textit{false negatives} in the full dataset, meaning that a FC-article actually fact-checks an original tweet but the article is viewed as an irrelevant one (i.e., 100\% precision but less than 100\% recall) because the FC-article was not embedded in a fact-checker's reply. For example, an original tweet is fact-checked by both a Snopes article and a Politifact one but only the Snopes article was embedded in the fact-checker's reply to the original tweet while the Politifact one was not included in the reply. If we build a model on the full dataset, the false negatives may mislead our model. To mitigate impact of this problem, we split the full dataset into two sub datasets called {Snopes} and {Politifact} datasets. The former one contains pairs where FC-articles are from {\textit{snopes.com}} and the later one contains pairs where FC-articles are from {\textit{politifact.com}}. Note, there still may be false negatives in each sub dataset since an original tweet may have multiple fake news stories fact-checked by different articles from the same fact-checking website but a fact-checker did not embed all of the articles in the reply. But the number of false negatives under this case would be smaller than those in the full dataset.
In Snopes dataset, we have 11,202 positive pairs made by 11,167 tweets and 1,703 FC-articles. In PolitiFact dataset, we have 2,037 positive pairs made by 2,026 tweets and 467 FC-articles. There are 102 overlapping tweets between the two datasets. 
The number of unique original posters is 8,277 and 1,482 in Snopes and Politifact respectively. On average, each original poster posted $\sim$1.35 tweets in our datasets. 
\section{Data Analysis}
\noindent\textbf{Topics of original tweets/queries.} Since the topic of an original tweet is related to the topic of a corresponding FC-article, we extracted topics of relevant FC-articles to understand the topical distribution of tweets. By analyzing each FC-article, top 5 topics of tweets in Snopes are as follows: Politics (42.3\%), Fauxtography (22.7\%), Junk News (8.1\%), Uncategorized (6.8\%), Quotes (4.8\%). For Politifact, tweets’ topics are mostly about politics due to its political mission. In conclusion, our datasets captured various topics. 

\noindent\textbf{Similarity of text in tweets and text in images.}
As we utilize text in images to enhance ranking performance, we seek to understand how similar text in tweets and text in images. For each query/tweet having text in its images, we transformed its text in tweet and its text in images into two vectors of TF-IDF values, and computed their cosine similarity. From all queries of a dataset, we computed mean cosine similarity. The mean similarity is 0.083 and 0.102 for Snopes and Politifact respectively, indicating that text in tweets is less similar to text in images. The number of tweets/queries containing text in images is 8,494 (76\%) and 1,742 (86\%) for Snopes and Politifact respectively.

\section{Experiments}
\label{sec:experiment}

\subsection{Neural Ranking Baselines}
We compare with 9 state-of-the-art neural ranking baselines, divided into 3 groups as follows: (1) multimodal retrieval methods including DVSH \cite{cao2016deep} and TranSearch \cite{guo2018multi}, (2) semantic matching models including ESIM \cite{chen2017enhanced} and NSMN \cite{nie2019combining}, and (3) relevance matching methods including MatchPyramid \cite{pang2016text}, KNRM \cite{xiong2017end}, ConvKNRM \cite{dai2018convolutional}, CoPACRR \cite{hui2018co} and DUET \cite{mitra2017learning}.  \noindent Please see the appendix for details of the baselines.
\begin{figure}[t]
	\vspace{-10pt}
	\centering
	\subfigure[Snopes]{
		\label{fig:snopes_base_retrieval}
		\includegraphics[trim=5 0 0 10,clip,width=0.46\linewidth,height=1.5in]{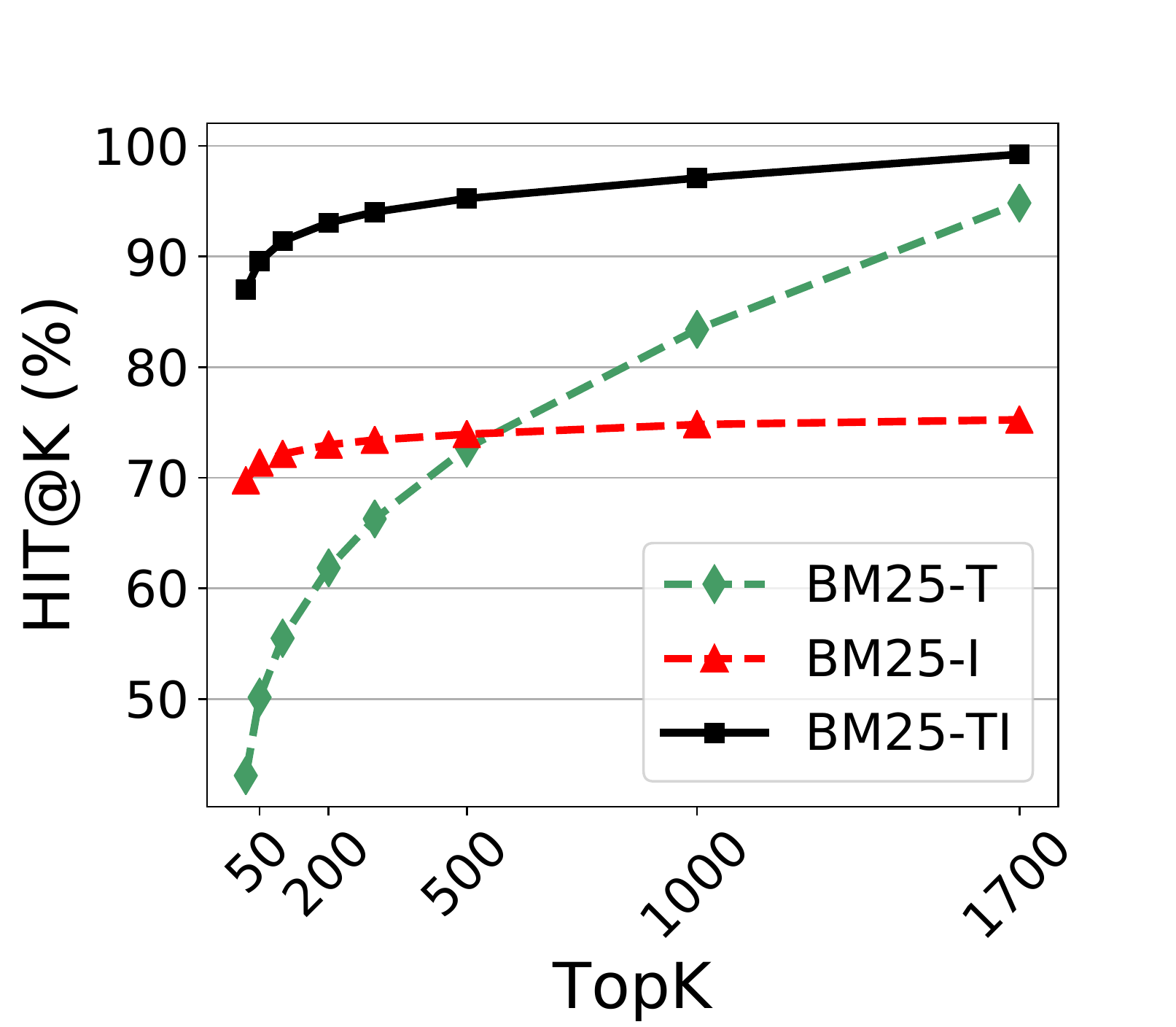}
	}
	\subfigure[Politifact]{
		\label{fig:politifact_base_retrieval}
		\includegraphics[trim=5 0 0 10,clip,width=0.46\linewidth,height=1.5in]{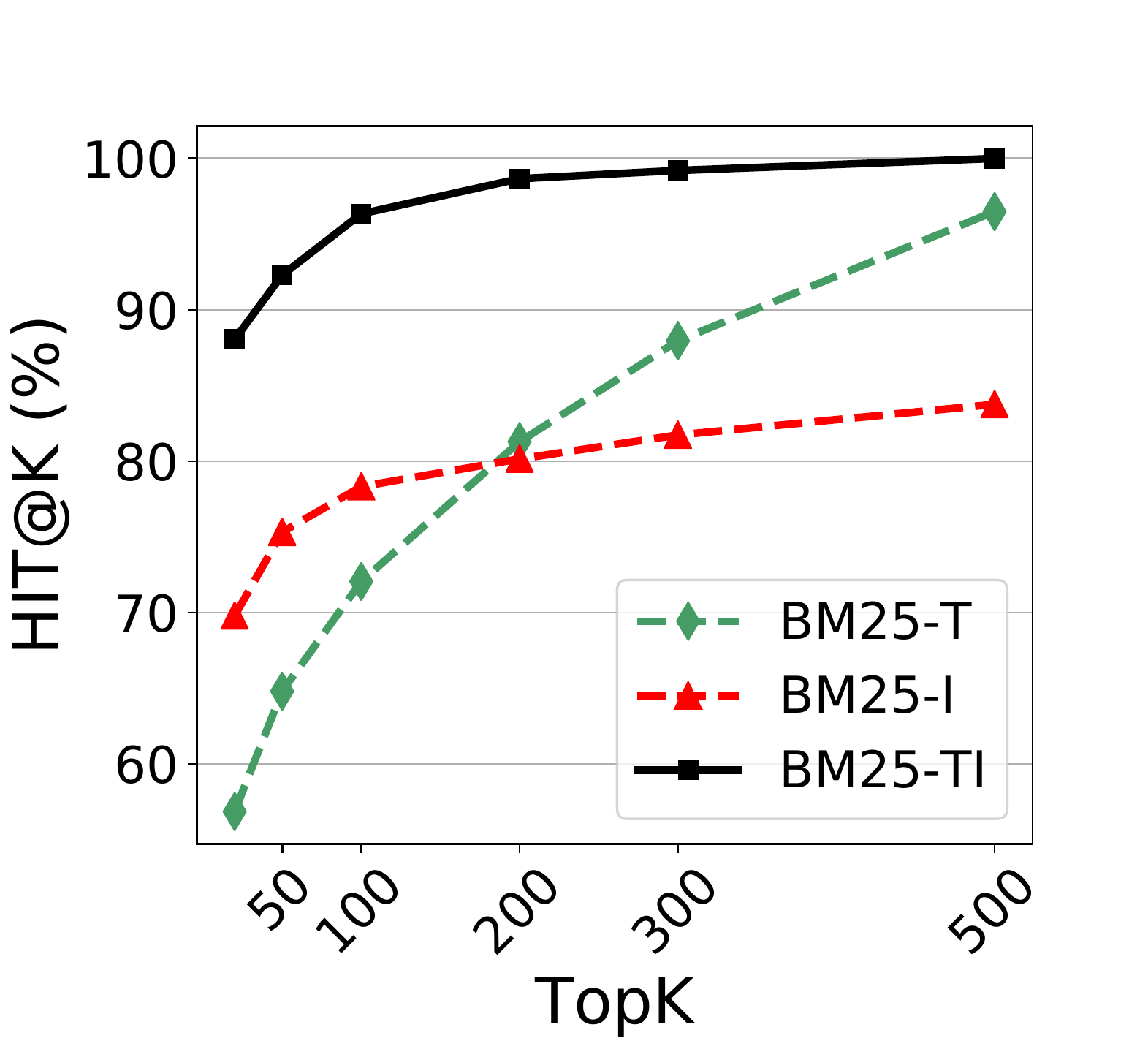}
	}
	\vspace{-10pt}
	\caption{Performance of basic retrieval methods}
	\label{fig:base_retrieval}
	\vspace{-15pt}
\end{figure}

\subsection{Experimental Design}
\label{sec:experimental_design}

\noindent\textbf{Evaluation Metrics. } We adopt NDCG@K \cite{xiong2017end} and HIT@K \cite{he2017neural} as evaluation metrics. We report mean HIT@K and NDCG@K where K $\in [1,3,5]$ based on all queries. Since over 99.5\% queries have only one relevant document, HIT@K is almost equal to Recall@K. Note, HIT@1 is equal to NDCG@1.

\noindent\textbf{Performance of the Basic Retrieval.} 
We test BM25 in three cases shown in Fig.~\ref{fig:base_retrieval}: (1) queries are tweets' text (BM25-T), (2) queries are text in tweets' images (BM25-I) and (3) queries are tweets' text + text in tweets' images (BM25-TI). 

In Fig.~\ref{fig:snopes_base_retrieval}, HIT@50 of BM25-T is only 50\% while BM25-I's HIT@50 is 70\%, suggesting that a lot of fake news appear in images. This is because tweets' text has at most 280 characters. Images are more attractive to online users and easier to convey fake news to them. When K is larger, BM25-I's HIT@K saturates quickly since only 76\% queries have text inside their images. Finally, BM25-TI is the best. Its HIT@50 is 89.6\%. 
Similar patterns appear in Politifact in Fig.~\ref{fig:politifact_base_retrieval}. 
With BM25-TI, HIT@50 is 94\%. From these results, we choose BM25-TI as the basic retrieval of our framework. 


\noindent\textbf{Split Datasets.} We need to choose value of K - the number of initial candidates for each query. If K is too small, initial candidates may not have relevant articles, leading to a meaningless re-ranking step. If K is too large, rerankers' running time may be high for online apps. We set K to 50 for both datasets. The number of queries, each of which has at least one relevant article in top 50 candidates, is 10,003 out of 11,167 for Snopes and 1,870 out of 2,026 for Politifact. Similar to \citet{thorne2018fever}, from these queries of each dataset, we randomly split them into train, validation and testing sets with ratio 80\%/10\%/10\% as shown in Table \ref{tbl:splited_datasets}. There are 1,164 and 156 leftover queries in Snopes and Politifact, respectively. Note, having leftover queries is a common issue for re-ranking based systems \cite{thorne2018fever}. Initial candidates output by BM25-TI are used by all neural ranking models.

\noindent\textbf{Testing Scenarios.} All models are tested in the re-ranking step with 2 scenarios (\textbf{SC1} and \textbf{SC2}). The main difference between them is whether to extract text from images of original tweets and FC-articles, and to incorporate the text with the other information (i.e., text and images of both the tweets and FC-articles). \textbf{SC1} and \textbf{SC2} are without text from images and with text from images, respectively.

\noindent\textbf{Experimental Settings.} For all baselines and our model, we use Adam optimizer \cite{kingma2014adam} with learning rate set to 0.001, and early stop training based on HIT@3 and NDCG@3 on a validation set with patience set to 10 epochs. Weight decay of L2-norm is 0.001. Batch size is 16. The number of negative documents sampled for each positive document during training is 3. Maxinum $|$words$|$ in queries is 50 for \textbf{SC1} and 100 for \textbf{SC2} respectively. Maximum $|$words$|$ in documents is 1,000. Vocab size $V$ in Snopes dataset is 25,932 in \textbf{SC1} and 40,670 in \textbf{SC2} respectively. Vocab size $V$ in Politifact dataset is 10,957 in \textbf{SC1} and 15,747 in \textbf{SC2} respectively. Maxinum $|$images$|$ in queries is 4. Maxinum $|$images$|$ in docs is 17. Images' shape is (224, 224, 3).

For our model, the number of  projection dimensions $P$ is chosen from $\{64, 128, 256, 512\}$. The number of output channels $F$ is chosen from $\{16,24\}$. The value of $k$ in kmax pooling is chosen from $\{16,32,48\}$. The number of CNNs $n$ is chosen from $\{1,2,3\}$.
Our model performs best on Snopes with $P$, $F$, $k$ and $n$ equal to 256, 16, 32 and 2, respectively. It performs best on PolitiFact with $P$, $F$, $k$ and $n$ equal to 256, 16, 48 and 3, respectively. 
We implement our model with PyTorch 0.4.1 and test it on a NVIDIA 1080 GTX GPU.

\begin{table}[t]
	\centering
	\caption{Split datasets}
	\vspace{-5pt}
	\resizebox{1.0\linewidth}{!} {	
		\begin{tabular}{l|lll|lll}
			\toprule[1pt]
			Datasets          & \multicolumn{3}{c|}{Snopes}                                                       & \multicolumn{3}{c}{Politifact}                                                  \\ \hline
			Items             & \multicolumn{1}{c}{Train} & \multicolumn{1}{c}{Valid} & \multicolumn{1}{c|}{Test} & \multicolumn{1}{c}{Train} & \multicolumn{1}{c}{Valid} & \multicolumn{1}{c}{Test} \\ \hline
			$|$Original Tweets$|$ & 8,002                     & 1,000                     & 1,001                     & 1,496                     & 187                       & 187                      \\
			$|$FC-Articles$|$     & 1,703                     & 1,697                     & 1,697                     & 467                       & 467                       & 467                      \\ \bottomrule[1pt]
		\end{tabular}
	}
	\label{tbl:splited_datasets}
	\vspace{-15pt}
\end{table}

\begin{table*}[t]
	\caption{Performance of our models and baselines when using images and text in tweets }
	\vspace{-5pt}
	\resizebox{1.0\linewidth}{!} {
		\begin{tabular}{cl|ccccc|ccccc}
			\toprule[1pt]
			\multicolumn{1}{c|}{\multirow{2}{*}{\begin{tabular}[c]{@{}c@{}}Ranking \\ Models Types\end{tabular}}}          & \multicolumn{1}{c|}{\multirow{2}{*}{\begin{tabular}[c]{@{}c@{}}Ranking\\ Models\end{tabular}}} & \multicolumn{5}{c|}{Snopes}                                                                                                                   & \multicolumn{5}{c}{Politifact}                                                                                                              \\ \cline{3-12}
			\multicolumn{1}{c|}{}                                                                                           & \multicolumn{1}{c|}{}                                                                           & \multicolumn{1}{l}{NDCG@1} & \multicolumn{1}{l}{NDCG@3} & \multicolumn{1}{l}{HIT@3} & \multicolumn{1}{l}{NDCG@5} & \multicolumn{1}{l|}{HIT@5} & \multicolumn{1}{l}{NDCG@1} & \multicolumn{1}{l}{NDCG@3} & \multicolumn{1}{l}{HIT@3} & \multicolumn{1}{l}{NDCG@5} & \multicolumn{1}{l}{HIT@5} \\ \hline
			\multicolumn{1}{c|}{Exact Matching}                                                                             & BM25-T                                                                                          & 0.20579                    & 0.27642                    & 0.32867                   & 0.30420                    & 0.39461                    & 0.18182                    & 0.29162                    & 0.37968                   & 0.31348                    & 0.43316                   \\ \hline
			\multicolumn{1}{c|}{\multirow{2}{*}{\begin{tabular}[c]{@{}c@{}}Multimodal \\ Retrieval (Group 1)\end{tabular}}} & DVSH-B                                                                                          & 0.38661                    & 0.51091                    & 0.60040                   & 0.54084                    & 0.67333                    & 0.26203                    & 0.33333                    & 0.38503                   & 0.36003                    & 0.44920                   \\
			\multicolumn{1}{c|}{}                                                                                           & TransSearch                                                                                     & 0.31668                    & 0.46081                    & 0.56444                   & 0.50062                    & 0.66034                    & 0.28342                    & 0.37925                    & 0.44920                   & 0.40040                    & 0.50267                   \\ \hline
			\multicolumn{1}{c|}{\multirow{2}{*}{\begin{tabular}[c]{@{}c@{}}Semantic Matching\\ (Group 2)\end{tabular}}}     & ESIM                                                                                            & 0.33367                    & 0.46608                    & 0.56444                   & 0.50372                    & 0.65534                    & 0.14973                    & 0.28722                    & 0.39037                   & 0.34871                    & 0.53476                   \\
			\multicolumn{1}{c|}{}                                                                                           & NSMN                                                                                            & 0.45754                    & 0.60097                    & 0.70330                   & 0.63220                    & 0.77822                    & 0.37968                    & 0.47718                    & 0.55080                   & 0.53128                    & 0.67914                   \\ \hline
			\multicolumn{1}{c|}{\multirow{5}{*}{\begin{tabular}[c]{@{}c@{}}Relevance Matching\\ (Group 3)\end{tabular}}}    & DUET                                                                                            & 0.36863                    & 0.48875                    & 0.57842                   & 0.52628                    & 0.66833                    & 0.29412                    & 0.41009                    & 0.49733                   & 0.43505                    & 0.55615                   \\
			\multicolumn{1}{c|}{}                                                                                           & MatchPyramid                                                                                    & 0.48052                    & 0.58523                    & 0.66034                   & 0.61565                    & 0.73327                    & 0.29412                    & 0.38903                    & 0.45455                   & 0.40812                    & 0.50267                   \\
			\multicolumn{1}{c|}{}                                                                                           & KNRM                                                                                            & 0.48951                    & 0.61081                    & 0.69730                   & 0.63686                    & 0.76124                    & 0.42246                    & 0.54935                    & 0.63636                   & 0.58456                    & 0.72193                   \\
			\multicolumn{1}{c|}{}                                                                                           & ConvKNRM                                                                                        & 0.52148                    & 0.63168                    & 0.70929                   & 0.65942                    & 0.77522                    & 0.45989                    & 0.57229                    & 0.65241                   & 0.62117                    & 0.77005                   \\
			\multicolumn{1}{c|}{}                                                                                           & CoPACRR                                                                                         & 0.53247                    & 0.64469                    & 0.72328                   & 0.67208                    & 0.78921                    & 0.45455                    & 0.59344                    & 0.69519                   & 0.62761                    & 0.77540                   \\ \hline
			\multicolumn{1}{c|}{\multirow{3}{*}{Ours}}                                                                      & CTM                                                                                             & 0.55744                    & 0.67555                    & 0.75624                   & 0.70156                    & 0.81918                    & 0.47059                    & 0.61669                    & 0.71658                   & 0.64292                    & 0.78075                   \\
			\multicolumn{1}{c|}{}                                                                                           & VMN                                                                                             & 0.68931                    & 0.73540                    & 0.76723                   & 0.75019                    & 0.80320                    & 0.24599                    & 0.26821                    & 0.31551                   & 0.28363                    & 0.35829                   \\
			\multicolumn{1}{c|}{}                                                                                           & MAN                                                                                             & \textbf{0.74326}           & \textbf{0.82197}           & \textbf{0.87712}          & \textbf{0.83447}           & \textbf{0.90609}           & \textbf{0.55080}           & \textbf{0.65435}           & \textbf{0.73262}          & \textbf{0.67644}           & \textbf{0.78610}          \\ \hline
			\multicolumn{2}{c|}{MAN vs. the best result of baselines}                                                                                                                                                                 & 39.59\%                    & 27.50\%                    & 21.27\%                   & 24.16\%                    & 14.81\%                    & 19.77\%                    & 10.26\%                    & 5.38\%                    & 7.78\%                     & 1.38\%                    \\ \bottomrule[1pt]
		\end{tabular}
	}
	\vspace{-10pt}
	\label{tbl:results_reranking_only_using_tweet_text}
\end{table*}

\begin{table*}[t]
	\caption{Performance of our models and baselines when using images, text in tweets and text in images }
	\vspace{-5pt}
	\resizebox{1.0\linewidth}{!} {
		\begin{tabular}{c|l|ccccc|ccccc}
			\toprule[1pt]
			\multirow{2}{*}{\begin{tabular}[c]{@{}c@{}}Ranking\\ Models Types\end{tabular}}            & \multicolumn{1}{c|}{\multirow{2}{*}{\begin{tabular}[c]{@{}c@{}}Ranking\\ Models\end{tabular}}} & \multicolumn{5}{c|}{Snopes}                                                                                                                   & \multicolumn{5}{c}{Politifact}                                                                                                              \\ \cline{3-12}
			& \multicolumn{1}{c|}{}                                                                          & \multicolumn{1}{l}{NDCG@1} & \multicolumn{1}{l}{NDCG@3} & \multicolumn{1}{l}{HIT@3} & \multicolumn{1}{l}{NDCG@5} & \multicolumn{1}{l|}{HIT@5} & \multicolumn{1}{l}{NDCG@1} & \multicolumn{1}{l}{NDCG@3} & \multicolumn{1}{l}{HIT@3} & \multicolumn{1}{l}{NDCG@5} & \multicolumn{1}{l}{HIT@5} \\ \hline
			Exact Matching                                                                             & BM25-TI                                                                                        & 0.63736                    & 0.69650                    & 0.73826                   & 0.71058                    & 0.77223                    & 0.27807                    & 0.34928                    & 0.40642                   & 0.38909                    & 0.50267                   \\ \hline
			\multirow{2}{*}{\begin{tabular}[c]{@{}c@{}}Multimodal \\ Retrieval (Group 1)\end{tabular}} & DVSH-B                                                                                         & 0.32667                    & 0.46849                    & 0.56843                   & 0.49640                    & 0.63636                    & 0.21925                    & 0.29335                    & 0.34759                   & 0.32626                    & 0.42246                   \\
			& TransSearch                                                                                    & 0.45854                    & 0.58410                    & 0.67433                   & 0.61832                    & 0.75724                    & 0.39572                    & 0.50878                    & 0.58824                   & 0.52397                    & 0.62567                   \\ \hline
			\multirow{2}{*}{\begin{tabular}[c]{@{}c@{}}Semantic Matching\\ (Group 2)\end{tabular}}     & ESIM                                                                                           & 0.61139                    & 0.70660                    & 0.77323                   & 0.72999                    & 0.83117                    & 0.33155                    & 0.44658                    & 0.52941                   & 0.48617                    & 0.62567                   \\
			& NSMN                                                                                           & 0.78821                    & 0.85732                    & 0.90809                   & 0.87148                    & 0.94106                    & 0.58824                    & 0.70002                    & 0.77540                   & 0.73500                    & 0.86096                   \\ \hline
			\multirow{5}{*}{\begin{tabular}[c]{@{}c@{}}Relevance Matching\\ (Group 3)\end{tabular}}    & DUET                                                                                           & 0.51848                    & 0.63605                    & 0.71928                   & 0.67075                    & 0.80220                    & 0.41711                    & 0.53087                    & 0.60963                   & 0.55757                    & 0.67380                   \\
			& MatchPyramid                                                                                   & 0.86513                    & 0.91150                    & 0.94406                   & 0.91791                    & 0.95904                    & 0.64171                    & 0.74872                    & 0.82353                   & 0.77702                    & 0.89305                   \\
			& KNRM                                                                                           & 0.84815                    & 0.89118                    & 0.92008                   & 0.90271                    & 0.94805                    & 0.65775                    & 0.75464                    & 0.82353                   & 0.77237                    & 0.86631                   \\
			& ConvKNRM                                                                                       & 0.85914                    & 0.90829                    & 0.94306                   & 0.91401                    & 0.95704                    & 0.66310                    & 0.79163                    & 0.88235                   & 0.80705                    & 0.91979                   \\
			& CoPACRR                                                                                        & 0.86913                    & 0.91166                    & 0.94006                   & 0.91851                    & 0.95604                    & 0.66845                    & 0.77419                    & 0.84492                   & 0.79191                    & 0.88770                   \\ \hline
			\multirow{3}{*}{Ours}                                                                      & CTM                                                                                            & 0.89910                    & 0.93191                    & 0.95504                   & 0.94008                    & 0.97502                    & 0.71123                    & 0.82512                    & 0.89840                   & 0.84331                    & 0.94118                   \\
			& MAN                                                                                            & 0.88412                    & 0.92563                    & 0.95604                   & 0.93238                    & 0.97203                    & 0.72193                    & 0.83104                    & 0.90374                   & 0.85313                    & \textbf{0.95722}          \\
			& MAN-A                                                                                          & \textbf{0.90909}           & \textbf{0.94204}           & \textbf{0.96503}          & \textbf{0.94892}           & \textbf{0.98202}           & \textbf{0.74332}           & \textbf{0.84905}           & \textbf{0.91979}          & \textbf{0.85987}           & 0.94652                   \\ \hline
			\multicolumn{2}{c|}{MAN-A vs. best result of baselines}                                                                                                                                        & 4.60\%                     & 3.33\%                     & 2.22\%                    & 3.31\%                     & 2.40\%                     & 11.20\%                    & 7.25\%                     & 4.24\%                    & 6.54\%                     & 2.91\%                    \\ \bottomrule[1pt]
		\end{tabular}
	}
	\vspace{-10pt}
	\label{tbl:results_reranking_using_text_inside_images}
\end{table*}


\subsection{Performance of Multimodal Attention Network and Variants}
\label{sec:man_results}
We also show results of MAN's variants as follows: (1) only using text (Eq.~\ref{eq:textual_features}) and (2) only using images (Eq.~\ref{eq:max_visual_consine_sim}). We call the former \emph{Contextual Text Matching} (CTM) and the later \emph{Visual Matching Network} (VMN). 
We show MAN's improvements wrt. the best result of baselines in each metric.

\noindent\textbf{SC1: Re-ranking using images and text in tweets.}
\label{sec:cman_sc1}
In Table \ref{tbl:results_reranking_only_using_tweet_text}, our CTM outperforms the best baselines, achieving maximum improvements of 4.7\% on NDCG@1. Our VMN amazingly outperforms text-based ranking baselines in Snopes perhaps because fauxtography is one of the most popular categories on Snopes \cite{friggeri2014rumor} while Politifact mainly fact-checks political claims. By using both text and images, 
our MAN shows an average increase of 17.2\% over the best baselines with the maximum improvement of 39.6\%.


\noindent\textbf{SC2: Re-ranking using images, tweets' text and images' text.}
We omit VMN from Table \ref{tbl:results_reranking_using_text_inside_images} since its results are same as Table \ref{tbl:results_reranking_only_using_tweet_text}. In Table \ref{tbl:results_reranking_using_text_inside_images}, both our MAN and CTM outperform baselines on two datasets. Interestingly, MAN has lower performance than CTM on Snopes while it has higher performance than CTM on Politifact. 
We suspect that the abundance of textual signals between original tweets and FC-articles in \textbf{SC2} unintentionally makes MAN tend to favor textual signals and neglect visual signals. 
To remedy this issue, we propose to augment training data in \textbf{SC2} with training data in \textbf{SC1} while keeping the same validation and testing set from \textbf{SC2}. Intuitively, the augmented training data may regularize MAN better \cite{yu2018qanet} by letting it observe both rich textual overlapping pairs in \textbf{SC2} and pairs with sparse textual signals in \textbf{SC1}. 
We name our model trained under the augmented training data as MAN-A. In Table \ref{tbl:results_reranking_using_text_inside_images}, 
MAN-A mitigates the above issue with an average increase of 4.8\% over the best baselines with the maximum improvement of 11.2\%. 
Text in images has a high impact on performance of CTM and MAN. In Table \ref{tbl:results_reranking_using_text_inside_images}, when using text in images to expand textual content of queries, performance of CTM and MAN increased by 17$\sim$34\% compared with their performances in Table \ref{tbl:results_reranking_only_using_tweet_text}.

From Tables \ref{tbl:results_reranking_only_using_tweet_text} and \ref{tbl:results_reranking_using_text_inside_images}, semantic matching models and multimodal baselines perform worse than relevance matching methods because the first two groups' goal is to compress whole queries and articles into dense vectors and measure their similarities. However, when compressing textual contents, some irrelevant information may be captured, leading to poor representations \cite{rao2019bridging}. 

In conclusion, our model MAN outperforms all baselines in both two testing scenarios .

\noindent\textbf{Experiments on the leftover original tweets (i.e., 1,164 tweets in Snopes and 156 tweets in Politifact).}
We further test benefits of using text and images on each leftover query where we rank its $x$ relevant articles against $50-x$ negative documents randomly sampled by following \citet{wan2016deep,wu2017sequential}. It means there are 50 FC-articles per query/tweet. Table \ref{tbl:additional_experiments} shows results of our best model MAN-A and best baselines in each group. As expected, MAN-A outperforms all the baselines due to sparse textual content in leftover queries.

\begin{table}[t]
	\caption{Ranking performances on leftover queries when using images, text in tweets and text in images}
	\vspace{-5pt}
	\resizebox{1.0\linewidth}{!} {
		\begin{tabular}{l|ccc|ccc}
			\toprule[1pt]
			\multicolumn{1}{c|}{\multirow{2}{*}{\begin{tabular}[c]{@{}c@{}}Ranking\\ Models\end{tabular}}} & \multicolumn{3}{c|}{Snopes}                                                          & \multicolumn{3}{c}{Politifact}                                                     \\ \cline{2-7}
			\multicolumn{1}{c|}{}                                                                           & \multicolumn{1}{l}{NDCG@1} & \multicolumn{1}{l}{NDCG@3} & \multicolumn{1}{l|}{HIT@3} & \multicolumn{1}{l}{NDCG@1} & \multicolumn{1}{l}{NDCG@3} & \multicolumn{1}{l}{HIT@3} \\ \hline
			TransSearch                                                                                     & 0.20361                    & 0.31856                    & 0.40292                    & 0.12821                    & 0.23542                    & 0.31410                   \\ \hline
			NSMN                                                                                            & 0.32646                    & 0.41123                    & 0.47595                    & 0.34615                    & 0.46871                    & 0.55769                   \\ \hline
			MatchPyramid                                                                                    & 0.26031                    & 0.33194                    & 0.38488                    & 0.28846                    & 0.34257                    & 0.37821                   \\
			ConvKNRM                                                                                        & 0.29124                    & 0.40280                    & 0.48282                    & 0.36538                    & 0.53479                    & 0.65385                   \\
			CoPACRR                                                                                         & 0.30928                    & 0.40748                    & 0.48024                    & 0.33333                    & 0.43887                    & 0.51923                   \\ \hline
			MAN-A                                                                                           & \textbf{0.58591}           & \textbf{0.68348}           & \textbf{0.75258}           & \textbf{0.51282}           & \textbf{0.64598}           & \textbf{0.73718}          \\ \hline
			Impr. MAN-A                                                                                     & 79.47\%                    & 66.20\%                    & 55.87\%                    & 40.35\%                    & 20.79\%                    & 12.74\%                   \\
			\bottomrule[1pt]
		\end{tabular}
	}
	\label{tbl:additional_experiments}
	\vspace{-5pt}
\end{table}

\begin{table}[t]
	\caption{Effects of contextual word embeddings}
	\vspace{-5pt}
	\resizebox{1.0\linewidth}{!} {
		\begin{tabular}{l|ccc|ccc}
			\toprule[1pt]
			\multicolumn{1}{c|}{\multirow{2}{*}{\begin{tabular}[c]{@{}c@{}}Ranking\\ Models\end{tabular}}} & \multicolumn{3}{c|}{Snopes}                                                          & \multicolumn{3}{c}{PolitiFact}                                                     \\ \cline{2-7}
			\multicolumn{1}{c|}{}                                                                           & \multicolumn{1}{l}{NDCG@1} & \multicolumn{1}{l}{NDCG@3} & \multicolumn{1}{l|}{HIT@3} & \multicolumn{1}{l}{NDCG@1} & \multicolumn{1}{l}{NDCG@3} & \multicolumn{1}{l}{HIT@3} \\ \hline
			Glove                                                                                           & 0.84216                    & 0.90017                    & 0.94106                    & 0.60428                    & 0.75713                    & 0.86096                   \\
			ELMo                                                                                            & 0.88511                    & 0.92865                    & \textbf{0.95804}           & 0.70588                    & 0.80080                    & 0.86631                   \\ \hline
			Glove+ELMo                                                                                      & \textbf{0.89910}           & \textbf{0.93191}           & 0.95504                    & \textbf{0.71123}           & \textbf{0.82512}           & \textbf{0.89840}          \\ \bottomrule[1pt]
		\end{tabular}
	}
	\vspace{-10pt}
	\label{tbl:effect_of_contextual_word_embeddings}
\end{table}

\begin{figure*}[t]
	\centering
	\subfigure[Matrix $\textbf{S}$ in Eq.~\ref{eq:cosine_sime_S}]{
		\label{fig:glove_cosine_sim_mat}
		\includegraphics[trim=5 0 0 10,clip,width=0.23\linewidth,height=1.5in]{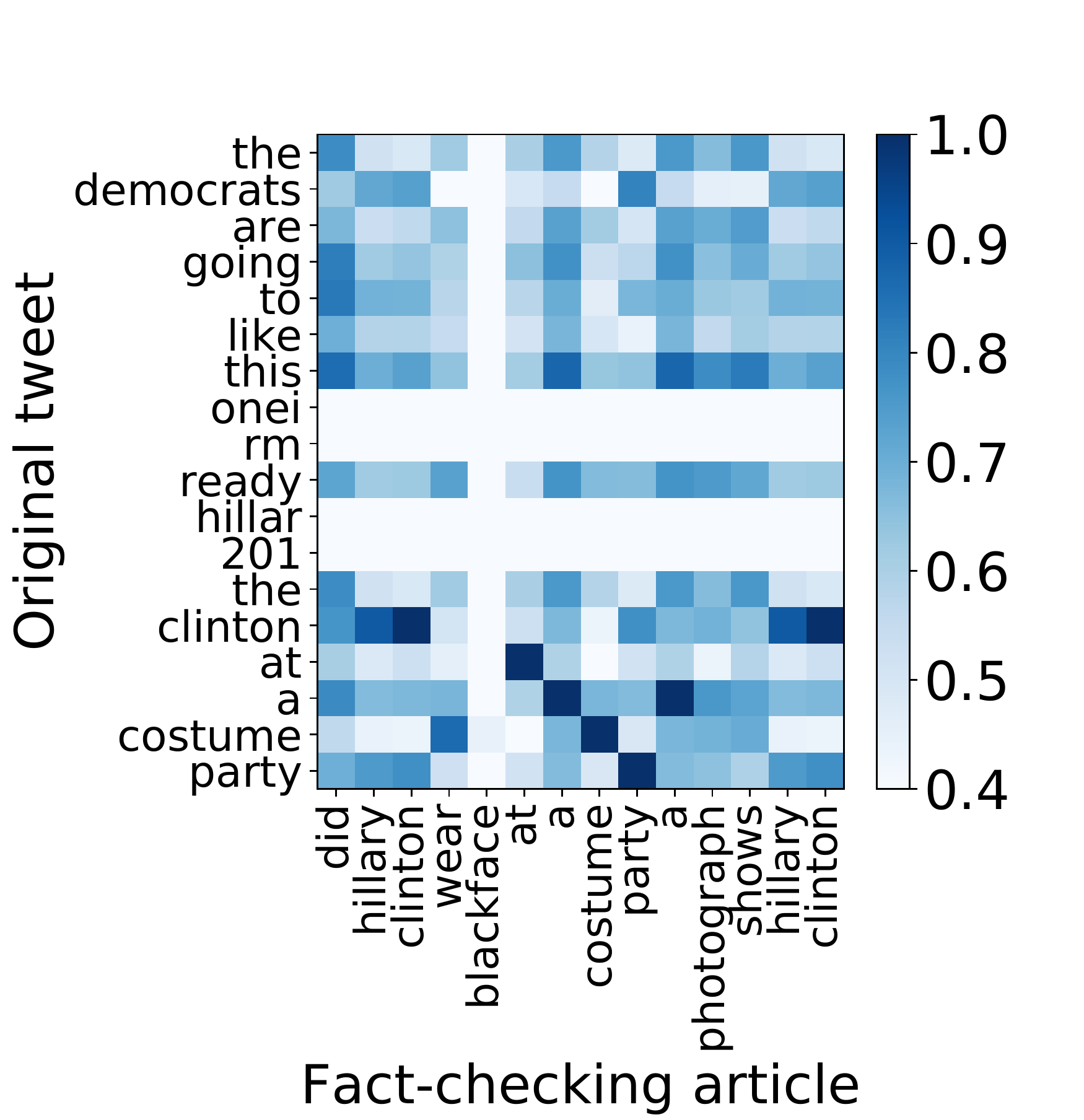}
	}
	\subfigure[Matrix $\textbf{G}$ in Eq.~\ref{eq:disim_cosine}]{
		\label{fig:gating_matrix}
		\includegraphics[trim=5 0 0 10,clip,width=0.23\linewidth,height=1.5in]{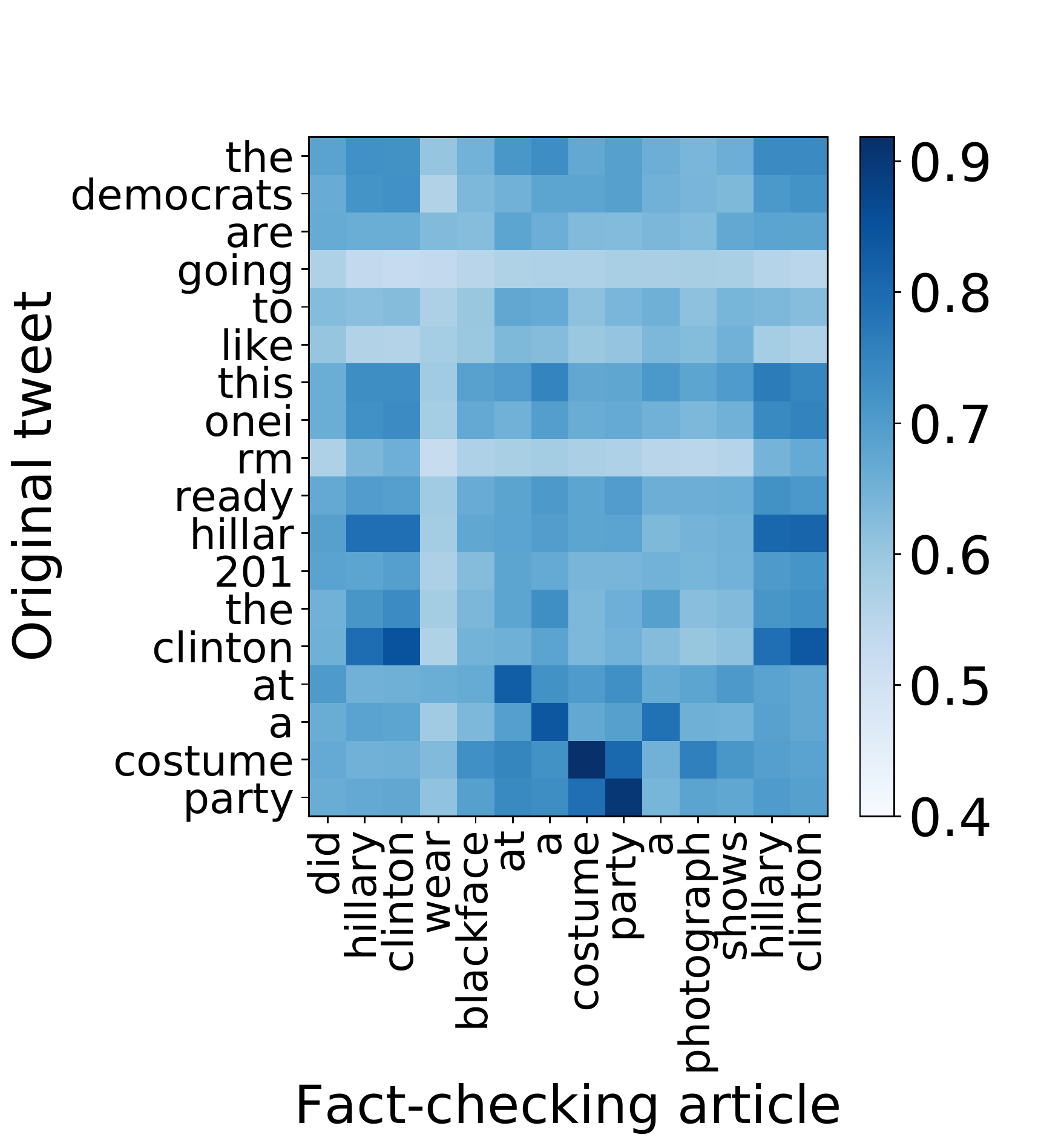}
	}
	\subfigure[Matrix $\textbf{A}$ in Eq.~\ref{eq:attended_matrix}]{
		\label{fig:attended_matrix}
		\includegraphics[trim=5 0 0 10,clip,width=0.23\linewidth,height=1.5in]{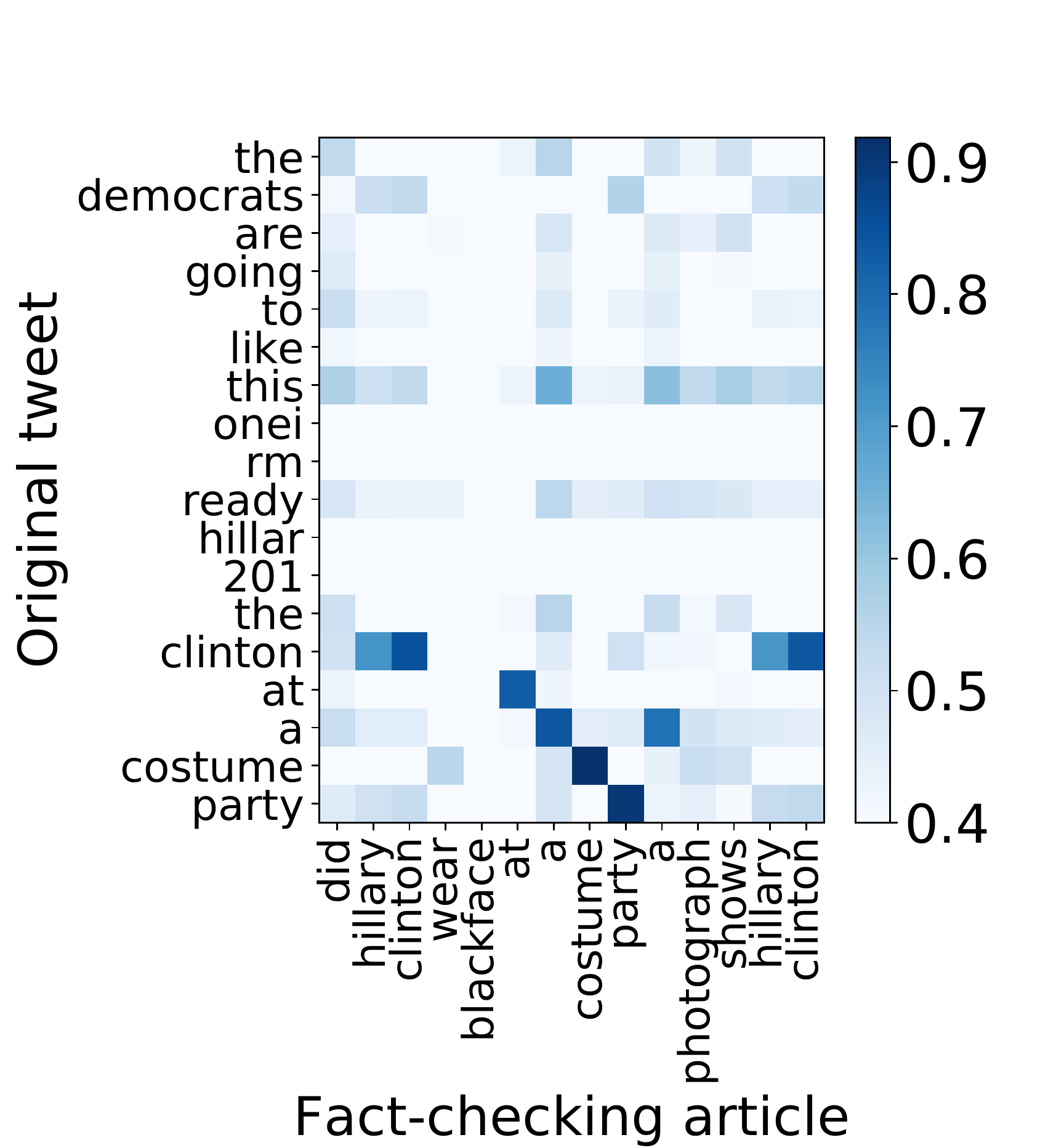}
	}
	\subfigure[Matrix $\textbf{C}$ in Eq.~\ref{eq:context_dot}]{
		\label{fig:elmo_cosine_sim_mat}
		\includegraphics[trim=5 0 0 10,clip,width=0.23\linewidth,height=1.5in]{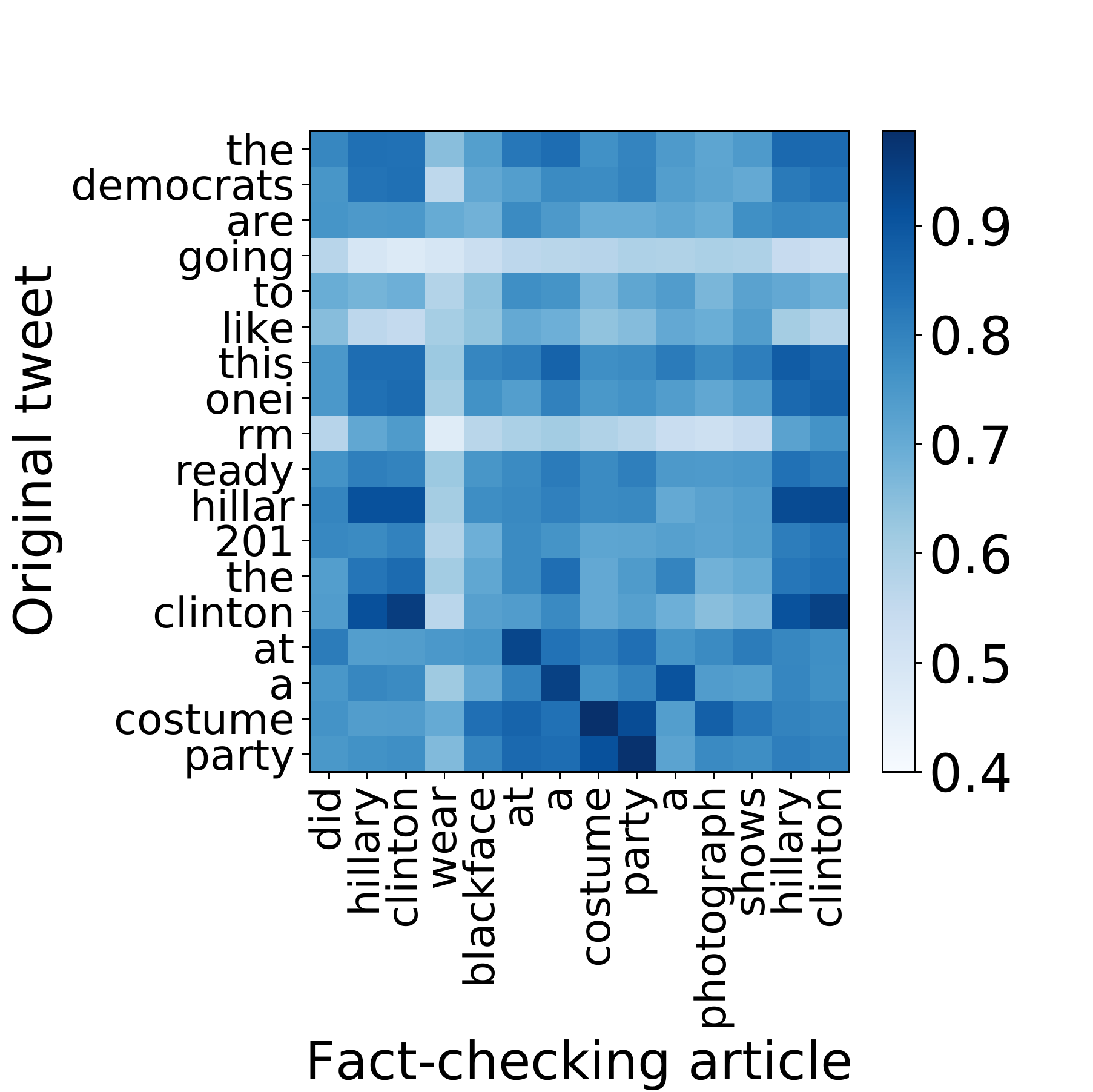}
	}
	\vspace{-5pt}
	\caption{Visualization of matrix \textbf{S}, matrix \textbf{G}, attended matrix \textbf{A} and matrix \textbf{C} (Best viewed in color)}
	\label{fig:viz_sim_matrices}
	\vspace{-10pt}
\end{figure*}

\subsection{Effect of Contextual Word Embeddings}
To understand effects of word embeddings on our model, we remove visual information and study re-ranking results of our model when (1) using only Glove embeddings, (2) using only contextual word embeddings from ELMo and (3) Glove+ELMo. In Table \ref{tbl:effect_of_contextual_word_embeddings}, when combining Glove and ELMo, we consistently achieve best NDCG in both datasets. 

\subsection{Case Studies}

\label{sec:case_studies}
\noindent\textbf{Qualitative comparison with the best baseline.} An example tweet is `You won't have to wait long \footnote{\url{https://bit.ly/3ngtsBK}}' embedded with a picture of an Antifa member beating a police officer. Clearly, the tweet’s text does not have any meaningful information while the image contains useful information. Given the tweet, the best baseline CoPACRR failed to find relevant FC-articles, whereas MAN ranked the correct FC-article \cite{AntiFaSnopes} in top-3 results.

\noindent\textbf{Visualization of interaction matrices and attended matrix}
Fig.~\ref{fig:viz_sim_matrices} visualizes matrices $\textbf{S}$, $\textbf{G}$, $\textbf{A}$ and $\textbf{C}$ in Eq.~\ref{eq:cosine_sime_S}, \ref{eq:disim_cosine}, \ref{eq:attended_matrix}, \ref{eq:context_dot} respectively of an original tweet and its FC-article from a testing set. Note, these matrices are learned by our model. In Fig.~\ref{fig:glove_cosine_sim_mat}, Glove embeddings help reveal overlapping phrases (e.g. \textit{at a costume party}, \textit{clinton}) but closeness of \textit{hillar} and \textit{hillary} is not well captured (i.e. sim(\textit{hillar}, \textit{hillary}) = 0.3). In contrast, sim(\textit{hillar}, \textit{hillary}) is 0.86 in Fig.~\ref{fig:elmo_cosine_sim_mat}, indicating quality of contextual word embeddings. 
The matrix $\textbf{G}$ in Fig.~\ref{fig:gating_matrix} has high values for key interactions (e.g. a list of values for \textit{at a costume party} is [0.56, 0.66, 0.68, 0.69]) in lieu of uniform values [1.0, 1.0, 1.0, 1.0] in matrix \textbf{S}). When combing matrix \textbf{S} and \textbf{G}, we have a sparse matrix \textbf{A} in Fig.~\ref{fig:attended_matrix} which pays more attention to key interactions (e.g. \textit{costume} and \textit{party}). 
In conclusion, the attention mechanism helps us capture key matching signals. 

\noindent\textbf{Impact of Searching for FC-Articles. }
We measure how much impact we can make on online users when correct FC-articles are retrieved (i.e. HIT@1 = 1). Totally, our best model, {MAN-A}, accurately finds FC-articles for 910 original tweets in test set of Snopes dataset. From these tweets, the total number of their retweets is 527,299 and total number of followers of the original posters who posted 910 original tweets is 233M. Roughly speaking, we can inform fact-checked information to millions of users. 
Security systems can prevent half million shares of fake news in those original tweets.

\section{Discussion}
\label{sec:discussion}

Since Snopes and Politifact are the most popular fact-checking sites,
building two models for them is an acceptable cost. 
When facing a real-life social media post, we run two trained models sequentially. If there is no found FC-article,
we can inform users that the post is unverified and suggest related pages from verified sites (e.g. governments' sites). When tweets do not have any images, we can use CTM which may find less relevant articles compared with MAN. However, CTM still performed better than the baselines as shown in Tables \ref{tbl:results_reranking_only_using_tweet_text} and \ref{tbl:results_reranking_using_text_inside_images}. We also built our best model (MAN-A) on the full dataset but observed some reduction in NDCG@1 and NDCG@3, but not HIT@3 compared with results of \textbf{SC2} on separate datasets maybe because of the \emph{false negatives} described in Section \ref{sec:datasets}. However, our model still outperformed the baselines. 

There are a few things our work could be improved. First, our basic retrieval BM25-TI does not consider images' similarities. To improve BM25-TI, we may combine images' similarities and BM25's score. We leave it as future work. 
Second, we create train/test data based on unique original tweets. Though there are no retweets and quotes, it is hard to completely ensure all queries's content are unique. However, our settings are applied to all models for fair comparisons. In addition, as shown in Fig.~\ref{fig:fake_news_example}, online users tend to re-post fake news. Therefore, it may be reasonable to have similar original tweets' content. Third, we tried to fine tune BERT but did not achieve good results perhaps because we did not have enough data. Interestingly, prior work \cite{shaar2020known} also had a similar observation when fine-tuning BERT. 

\section{Conclusions}
In this paper, we propose a novel method to alleviate the spread of fake news. By searching for FC-articles and incorporating fact-checked information into social media posts, we can warn users about fake news and discourage them from spreading misinformation.
Our framework uses text and images to search for FC-articles, achieving an average increase of 4.8\% over best baselines with the maximum improvement of 11.2\%. Complementary to fake news detection methods, our method proactively scales up verified content on social media.

Our framework can be used for other multimodal retrieval tasks (e.g. searching for verified sites as we suggested in the previous section).

\section*{Acknowledgment}
This work was supported in part by NSF grant CNS-1755536, AWS Cloud Credits for Research, and Google Cloud. Any opinions, findings and conclusions or recommendations expressed in this material are the author(s) and do not necessarily reflect those of the sponsors.
\newpage
\newpage

\bibliographystyle{acl_natbib}
\bibliography{www}

\clearpage
\appendix



\begin{figure*}[h]
	\centering
	\includegraphics[trim=0cm 0cm 4cm 0cm,clip,width=1\linewidth,height=3in]{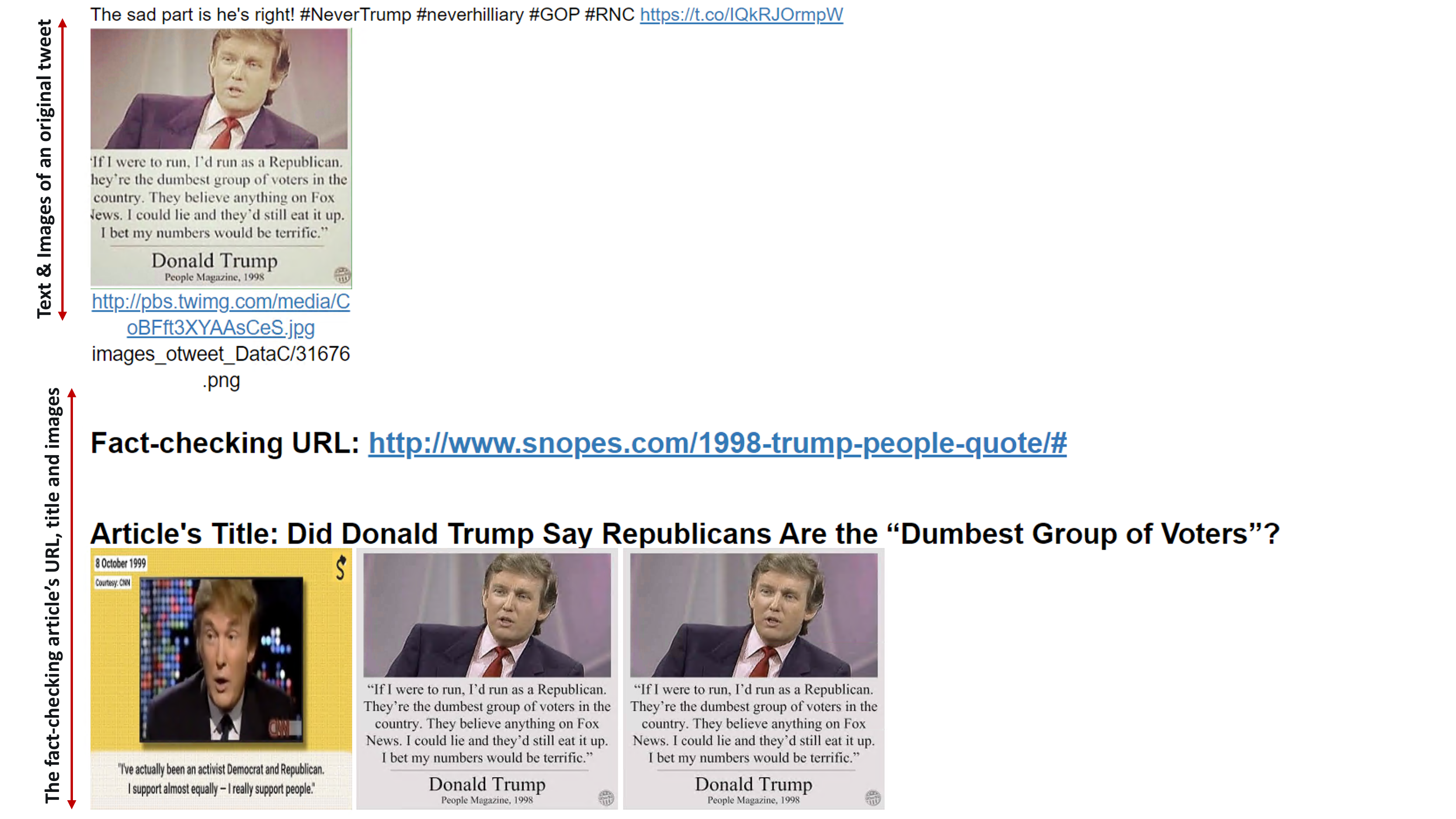}
	\caption{Labeling UI}
	\label{fig:labeling_ui}
\end{figure*}
\section{Appendix for the Reproducibility}

\subsection{Labeling UI}
We developed a labeling UI as shown in Fig.~\ref{fig:labeling_ui} to support labelers to quickly explore linked articles. It includes text and images of original tweets as well as text and images of a FC-article.


\subsection{Descriptions and Hyperparameters of the Baselines}

\noindent\textbf{Multimodal Retrieval Models.} DVSH \cite{cao2016deep} accepts a pair of a multimodal query and a multimodal article, and outputs similarity score. It uses cosine max-margin loss. We also tried to compare with DVSH by using hashcode of queries' text to match articles' images and vice versa. However, DVSH did not perform well perhaps because queries' text and documents' images may be not semantically similar. We implemented DVSH by ourselves because there is no publicly downloadable code. We set its hidden size to 300 and used AlexNet to extract visual features by following \cite{cao2016deep}.

TranSearch \cite{guo2018multi} learns representations of queries by using queries' text and representations of documents by using text and images of the documents. 
For TranSearch, we omitted the pretraining step because our datasets do not have \textit{also\_viewed} or \textit{buy\_after\_viewing} information. VGG19 was used to extract visual features by following \cite{guo2018multi}. We used the publicly accessible TranSearch implementation. 


\noindent\textbf{Semantic Matching Models.} We compare with ESIM \cite{chen2017enhanced} and NSMN \cite{nie2019combining}. Both models utilize BiLSTM encoders to learn contextual representations and measure similarity between queries and documents. NSMN also uses contextual word embeddings from ELMo with skip connections for better performance.
The pretrained ELMo\footnote{https://allennlp.org/elmo} with 93.6M parameters was used for {NSMN} and our proposed models. Its hidden size was 4,096, and output size was 512 with using 2 highway layers.

In ESIM, a hidden size was set to 300. In NSMN, we set its hidden size to 100 to all BiLSTM layers. We also tried to set its hidden size to $\{200, 300\}$ but we got out-of-memory error on our GPU because NSMN is memory-intensive due to concatenation of word embeddings, contextual embeddings from ELMo and multiple BiLSTM layers on our documents with 1,000 tokens.

\noindent\textbf{Relevance Matching Models.} We compare with several state-of-the-art models in this category. 
MatchPyramid \cite{pang2016text} uses CNN to capture spatial patterns. KNRM \cite{xiong2017end} and ConvKNRM \cite{dai2018convolutional} use RBF kernel to pool n-gram matching signals. CoPACRR \cite{hui2018co} uses similarities between queries' representations and context-aware representations of words in documents to attend to matching signals. DUET \cite{mitra2017learning} unifies semantic and relevance matching signals into one model.

Implementation of MatchPyramid, KNRM, ConvKNRM, and DUET was obtained from MatchZoo \cite{guo2019matchzoo}. In MatchPyramid, we used default setting of MatchZoo. The number of kernels of KNRM and ConvKNRM was chosen from $\{7, 9, 11\}$. In ConvKNRM, we set $|$filters$|$ to 300 like the word embeddings' dimension size. n-gram was chosen from $\{1, 2, 3\}$. In DUET, we followed the same architecture proposed in \cite{mitra2017learning}.

In CoPACRR, the number of CNN layers was chosen from $\{2, 3, 4\}$, the number of filters was chosen from $\{5, 10, 15\}$, top k largest matching signals $ns$ was chosen from \{5, 10, 15\}. The number of segments to conduct k-max pooling $cpos$ was chosen from \{4, 5, 6\} and context windows were chosen from $\{5, 9, 11\}$. We used the publicly accessible CoPACRR implementation.

\end{document}